\documentclass[11pt,a4paper]{article}
\usepackage{jinstpub}

\usepackage{amsmath}
\usepackage{graphicx}
\usepackage{xcolor}
\usepackage{natbib}
\usepackage{lineno}

\title{Light yield non-proportionality of inorganic crystals and its effect on cosmic-ray measurements}

\author[a,b]{O. Adriani}

\author[a,b]{E. Berti}

\author[a]{P. Betti}

\author[m,l]{G. Bigongiari}

\author[a,b]{L. Bonechi}

\author[a,b]{M. Bongi}

 \author[b]{S. Bottai}

\author[m,l]{P. Brogi}

\author[n,b]{G. Castellini}

\author[m,l]{C. Checchia}

\author[a,b]{R. D'Alessandro}

\author[b]{S. Detti}

\author[r,b]{N. Finetti}

\author[m,l]{P. Maestro}

\author[m,l]{P.S. Marrocchesi}

\author[a,b]{N. Mori}

\author[a,b]{M. Olmi}

\author[a,b]{L. Pacini}

\author[b]{P. Papini\footnote{Corresponding author papini@fi.infn.it} }

\author[a]{C. Poggiali}%

\author[n,b]{S. Ricciarini}

\author[a]{P. Spillantini}

\author[b]{O. Starodubtsev}

\author[m,l]{F. Stolzi}

\author[a,b]{A. Tiberio}

\author[b]{E. Vannuccini }

\affiliation[a]{Department of Physics and Astronomy, University of Florence, 
via G. Sansone 1, I-50019 Sesto Fiorentino (Firenze), Italy}
\affiliation [b]{INFN Firenze, 
via B. Rossi 1, I-50019 Sesto Fiorentino (Firenze), Italy}

\affiliation [m]{Department of Physical Sciences, Earth and Environment, University of Siena, 
I-53100 Siena, Italy}
\affiliation [l]{INFN Pisa, 
Largo B. Pontecorvo, 3 – 56127 Pisa, Italy}
\affiliation [n]{IFAC (CNR), 
via Madonna del Piano 10, I-50019 Sesto Fiorentino (Firenze), Italy}
\affiliation [r]{Department of Physical and Chemical Sciences, University of L'Aquila, 
Via Vetoio, Coppito, 67100 L'Aquila, Italy}

\emailAdd{papini@fi.infn.it}

\abstract{
The multi-TeV energy region  of the cosmic-ray spectra has been recently explored  by direct detection experiments that used calorimetric techniques to measure the energy of the cosmic particles.
Interesting spectral features have been observed in both all-electron and nuclei spectra. 
However, the interpretation of the results is compromised by the disagreements between the data obtained from the various experiments, that are not reconcilable with the quoted experimental uncertainties.
Understanding the reason for the discrepancy among the measurements is of fundamental importance in view of the forthcoming high-energy cosmic-ray experiments planned for space, as well as for the correct interpretation of the available results.

The purpose of this work is to investigate the possibility that a systematic effect may derive from the non-proportionality of the light response of inorganic crystals, typically used in high-energy calorimetry due to their  excellent energy-resolution performance. 
The main reason for the non-proportionality of the crystals is that  scintillation light yield depends on ionisation density. 
Experimental data obtained with ion beams were used to characterize the light response of various scintillator materials. 
The obtained luminous efficiencies were used as input of a Monte Carlo simulation to perform a comparative study of the effect of the light-yield non-proportionality on the detection of high-energy electromagnetic and hadronic showers.  
The result of this study indicates that, if the calorimeter response is calibrated by using the energy deposit of minimum ionizing particles, the measured shower energy might be affected by a significant systematic shift, at the level of few percent, whose sign and magnitude depend specifically on the type of scintillator material used. 

}


\begin{document}

\maketitle

\flushbottom

\section{Introduction}
\label{sec:intro}

Most of the recent direct cosmic-ray measurements at high energy (above the TeV region) are performed by means of calorimetric techniques to measure the energy of the cosmic particles. 

The light component of cosmic rays, the so-called all-electron spectrum (electrons plus positrons components), has been measured up to several TeV and at about $\sim$ 1 TeV a break to a softer spectrum has been reported by atmospheric indirect measurements (HESS \cite{HESS1} and VERITAS \cite{VERITAS1} experiments). This structure in the all-electron spectrum has been recently confirmed by two calorimetric experiments in space, DAMPE \cite{DAMPE1} and CALET \cite{CALET1} which made the first direct measurements at these energies. 

\begin{figure}[htb]
\begin{center}
\includegraphics[width=13.5cm]{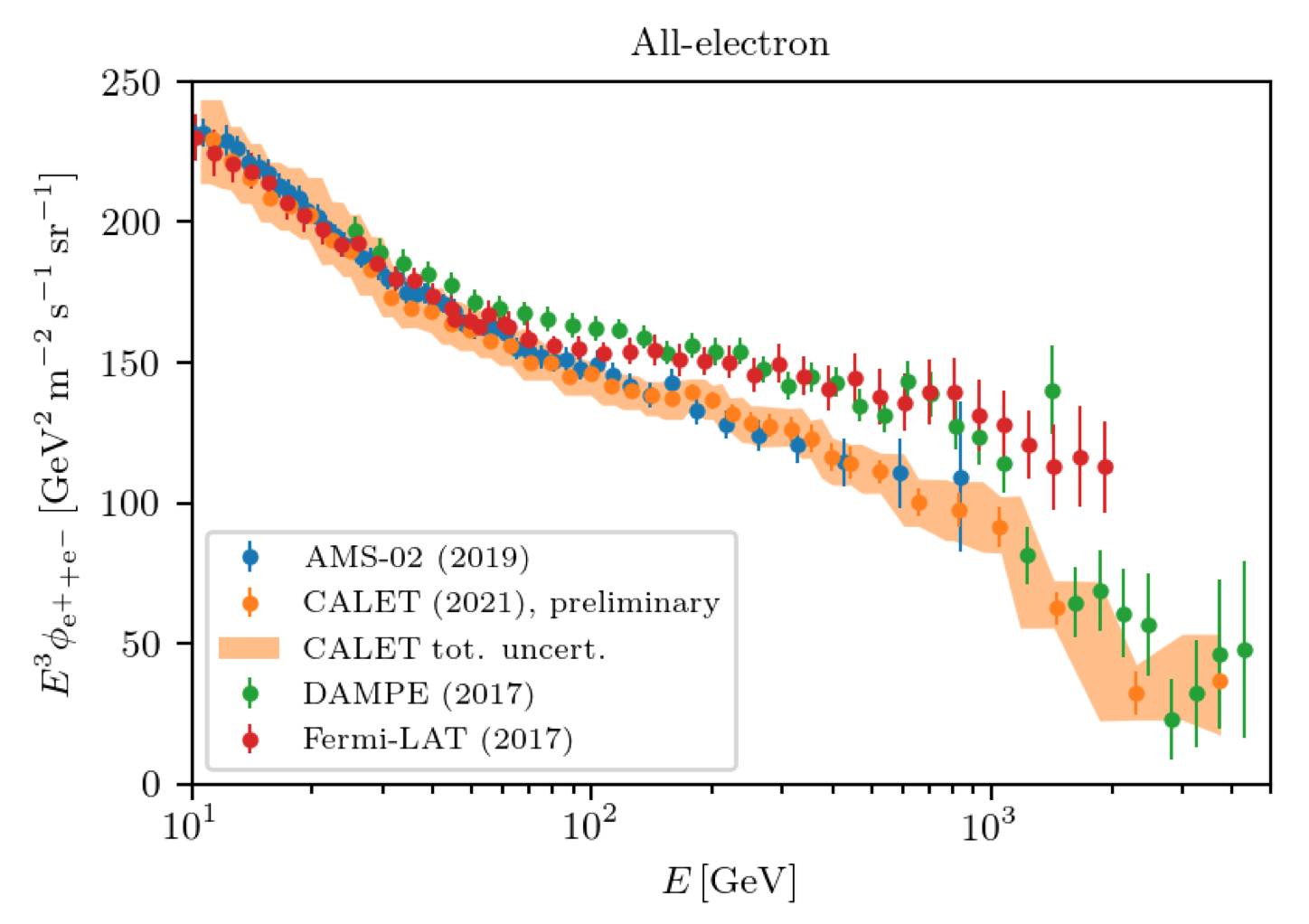}
\end{center}
\caption{Compilation of all-electron fluxes from direct measurements taken from the 37th International Cosmic Ray Conference \cite{ICRC2021}.
\label{fig:electronpositron}}
\end{figure}

The knowledge of the cosmic-ray spectra at these high energies is particularly relevant in connection with the positron excess evidence over pure secondary production and to model cosmic-ray production with a single source or by means of stochastic population of sources in the Galaxy. However, the spectrum interpretation is compromised by the disagreements between the data obtained from the various experiments. The differences are evident when we compare the data even at lower energies. Below 1 TeV there are also measurements of experiments based on a magnetic spectrometer, such as AMS-02 \cite{AMS1}, in addition to the FERMI calorimetric experiment \cite{FERMI1}. In Figure \ref{fig:electronpositron} a recent compilation of the all-electron spectrum, taken from the latest International Cosmic Rays Conference, is shown \cite{ICRC2021}. Above $\sim$30 GeV a disagreement appears in the spectra up to $\sim$40\% that is not reconcilable with the various quoted experimental uncertainties. At high energy the comparison in the measurements of electron and positron spectra carried out with calorimetric techniques is particularly worrisome because, in principle, they have an accuracy in the energy measurement of a few percent, much better than that obtainable with magnetic spectrometers.

Given this situation, it becomes urgent to be able to understand the reason for the discrepancies both for a correct interpretation of the data and for designing future experiments in space with better reliability.

For what concern cosmic nuclei measurements, space experiments have stringent weight limits and this typically leads to calorimetric depth no more than some interaction lengths ($\lambda_I$). Therefore, direct calorimetric measurements of cosmic-ray nuclei intrinsically have a worse precision due to limited energy resolution. Despite this, even when we compare measurements on the nuclei carried out with calorimeters sometimes we find discrepancies that cannot be attributed only to the estimated uncertainties. For instance the CALET experiment presented the recent measurements of carbon, oxygen and iron cosmic fluxes between $\sim$ 10 GeV/n and $\sim$ 2 TeV/n \cite{CALET2,CALET3}. While the spectral shape of these fluxes are basically in agreement with the AMS-02 results they differ in the absolute normalisation of about 27\%. This fact is particularly puzzling given the good agreement between the CALET and the AMS-02 measurements of all-electron flux (see Figure~\ref{fig:electronpositron}). It is clear that an effort is needed to understand the reason of these unknown systematic effects.

The measurements based on the calorimetric instruments are typically made with inorganic scintillating crystals and a possible systematic effect is due to the non-proportional light response of the crystals \cite{Moses1}. The main purpose of this article is to investigate this possibility. In Table~\ref{tab:experiments} a list of space experiments based on a calorimetric approach is shown with the main characteristics of the calorimeter materials and depths.

In Section \ref{sec:scintillation} the theoretical model to reproduce the light response of crystals is described. In the same section the technique employed to characterise the light response curve by means of data collected with particle beams is presented. In Section \ref{sec:beamtest} the results of a beam test are shown. The obtained crystal characterizations are used to derive the possible systematic effect on cosmic-ray measurements and the results are discussed in Section~\ref{sec:showers}. Conclusions are drawn in Section \ref{sec:conclusion}. 

\begin{table}[htb]
\begin{center}
\begin{tabular}{l|c|c|c|l}
Experiment & Material &  Electromagnetic & Hadronic & Launch year \\ 
 &   & depth ($X_{0}$) & depth ($\lambda_{I}$) &  \\ \hline
 CALET~\cite{TORII20192531} & PWO      &  27 & 1.2 & 2015 \\
 DAMPE~\cite{CHANG20176} & BGO      &  32 & 1.6 & 2015 \\
 FERMI~\cite{Atwood_2009} & CsI(Tl)  & 8.6  & 0.4 & 2008 \\
 HERD~\cite{Gargano:2021Q4}  & LYSO     & 55  & 3.0 & 2027 (expected) \\
\hline
\end{tabular}
\end{center}
\caption{List of space cosmic-ray experiments based on calorimetric instruments, made by inorganic scintillators, and the main characteristics of the calorimeters. 
\label{tab:experiments}}
\end{table}

\section{ Non-proportional light response}
\label{sec:scintillation}

The term "non-proportional" means the fact that the light response of a scintillating material is not exactly proportional to the energy released by a crossing particle, with the consequence that a detector response depends on both the species and energy of the particle that interacts with the scintillator.

It is known \cite{Murray1} that the main reason for the non-proportionality of the crystals is that the scintillation light yield depends on the ionisation density. When a charged particle passes through a scintillating material inevitably the ionisation density is non-uniform due to different cascade processes involved in the ionisation (elastic collisions, photoelectric or Compton interactions, delta-rays and fluorescent X-rays, Auger-like processes, etc.) leading to a total effect of non-proportionality of the light response.

In order to investigate this effect on particle detectors two ingredients are needed: a model to parametrise the light yield response as a function of the ionisation density (or other causes) and a calculation method to determine the distribution of the ionisation density when a particle crosses the detector. 

\subsection{Minimalist Approach}

The general picture was originally proposed in 1961 \cite{Murray1}. When a charged particle interacts in the material the ionisation process is usually modelled as a straight tube with a diameter of about 3 nm around the track in which three kinds of carriers are produced: excited states of single electrons, holes and "excitons", which are  bound pairs of particles consisting of an electron and a hole. The partition of the ionisation among the three kinds of carriers depends on both the scintillator material and the ionisation density.   

More recently, as described in Moses et al. \cite{Moses2}, three kinds of models are usually used to describe the scintillator non-proportionality. The first is the "minimalist approach" in which the minimum number of effects are assumed to be the most important ones with some simplified assumptions. The "kinetic model" considers also the time evolution of the ionisation density and the "diffusion model" considers also the carrier mobility. For our purpose, as described in the following, the minimalist approach is able to reproduce and describe the experimental data with sufficient accuracy. In this model a simplified assumption is taken: considering that neutral carriers are much more mobile than charged ones it is assumed that only excitons are able to transfer energy to luminescent centres, and therefore charged carriers (electrons and holes) recombine non-radiatively.   

To write down the luminous efficiency formula two phenomena are considered. 
\begin{enumerate}
  \item At high excitation density the quenching (or Birks) effect is dominant, where an Auger-like mechanism involves two excitons. When two excitons collide, at the end of the interaction, part of their energy is transferred  non-radiatively to phonons. Only the remaining part, taken way by a single exciton, is available to produce luminescence. This is the mechanism described by Birks to model quenching \cite{Bircks1} and the rate at witch it occurs is proportional to the square of the exciton density. The original formula to describe the quenching effect provides the relative luminous Birks efficiency $L_B$ as function  of the linear ionisation density $dE/dx$ as
\begin{equation}
  L_B = \frac{1}{1+B \times \frac{dE}{dx}}
\end{equation}
where $B$ is the Birk's parameter. Assuming a division of the energy deposition into cylindrical "core" and "halo" regions surrounding the particle trajectory, a modification of the Birks formula has been proposed in \cite{Tarle1}. The carriers in the core region, where the ionisation density is higher, undergo the quenching effect, whereas in the halo region no saturation phenomenon is assumed. Let  $\eta_H$ be the fraction of carriers escaped to the halo region,  the relative luminous Birks efficiency formula is then modified as
\begin{equation}
  L_B' = \frac{1-\eta_H}{1+B(1-\eta_H) \times \frac{dE}{dx}} + \eta_H
\label{BirksMod}
\end{equation}
\item At low excitation density another phenomenon can be dominant. Let $\eta_{e/h}$ be the fraction of initial electrons and holes that do not form excitons. They are physically separated and only electrons and holes that are closer than the Onsager radius can combine to form excitons and therefore luminesce~\cite{Payne2,Moses2}. The amount of this effect depends on the carrier density and therefore on the $dE/dx$. The relative luminous Onsager efficiency $L_O$ can be written as  
\begin{equation}
  L_O = 1 - \eta_{e/h} \exp\left( - \frac{(dE/dx)}{(dE/dx)_{O}} \right)
\end{equation}
where $(dE/dx)_{O}$ is the strength of the Onsager term.
\end{enumerate}
Combining the modified Birks and Onsager mechanisms at high and low ionisation density respectively, the final expression for the relative luminous efficiency is:
\begin{equation}
  L = \left[ 1 - \eta_{e/h} \exp\left( - \frac{(dE/dx)}{(dE/dx)_{O}} \right) \right] \times \left[ \frac{1-\eta_H}{1+B(1-\eta_H)\times\frac{dE}{dx}} + \eta_H \right]. \label{Model}
\end{equation}
This expression with four parameters ($\eta_{e/h}$, $(dE/dx)_{O}$, $\eta_H$, and $B$) has been used in this work. It is a modification of the expression used in \cite{Payne2} and \cite{Moses2} in which a parameter was added (according to the modified Birks formula \ref{BirksMod}) to better describe the experimental data up to an ionisation density of the order of 2 GeV/cm as shown in the Section \ref{sec:beamtest}. 
Later in the paper it will be illustrated how the terms of the equation vary for some reference scintillator materials (see e.g. Figure \ref{fig:functions}).

\subsection{Monte Carlo simulation}

When a particle enters in a scintillator material many processes take place which produce ionising particles at different energies. Each one of the produced charged particles ionises the material with an ionisation density that depends on the energy and type of particle itself according to the Bethe-Bloch formula and statistical fluctuations. 

A simulation code has been developed based on the FLUKA tool \cite{FLUKA} in order to describe the particle interactions with the scintillator material. At the end of the simulated tracking processes a distribution of ionisation density is produced on an event-by-event basis. The minimum energy threshold for particle tracking has been reduced as much as possible compatibly with the possibilities of the FLUKA code, i.e. 1 keV for electrons and 100 eV for photons. Furthermore, particular care has been taken in activating  all the physical processes that can contribute to the amount of ionisation. 

\begin{figure}[htb]
\begin{center}
\includegraphics[width=13.5cm]{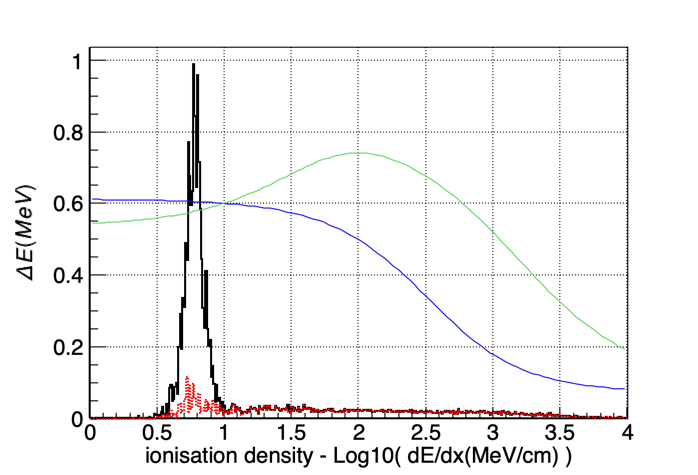}
\end{center}
\caption{Simulated ionisation density distribution for an Hydrogen nucleus at minimum ionization  crossing a 2-cm thick LYSO layer. The black histogram refers to the total energy loss whereas the red histogram refers to the contribution of $\delta$-rays only. Curves refer to the typical relative luminous efficiency for silicate (blue) and alkali (green) materials normalized to arbitrary units (the curve parameters are those of LYSO and CsI(Tl) in Table \ref{tab:fit}). 
\label{fig:MIPexample}}
\end{figure}

In Figure \ref{fig:MIPexample} an example is shown of the output relative to a minimum ionising particle that crosses a thin layer of material 
(in particular, a Hydrogen nucleus that crosses 2~cm of LYSO).
The code provides for every $i$-th bin of ionisation density $\left(\frac{dE}{dx}\right)_i$ the amount of the energy released in the scintillator $\Delta E_i$. The used binning size is uniform in log
scale and the total detectable energy loss by the particle in the material is
\begin{equation}
  E_{rel} = \sum_i \Delta E_i.
\end{equation}

The black histogram is the total energy loss distribution as a function of the ionisation density whereas the red histogram refers to the contribution of the delta-rays only. The ionisation due purely to the primary nucleus track is centred around 6 MeV/cm while the delta-ray ionisation spreads out in a large queue of more than 3 orders of magnitude with a contribution to the total energy loss of about 35\%. In the same figure, two typical examples of relative luminous efficiency $L\left( \frac{dE}{dx}\right)$ in arbitrary units are also shown: the blue curve has the typical functional shape for the silicate materials (like LYSO or GSO) while the green curve for the alkali materials (like CsI(Tl) or NaI) \cite{Payne}. 

The total light signal $S_L$ in arbitrary units is given by the energy loss weighted with the relative luminous efficiency:
\begin{equation}
  S_L = \sum_i \Delta E_i \times L_i.
\end{equation}

Different materials have different behaviours, in particular in case of silicate crystals the delta-ray contribution 
weighs comparatively less than for alkali materials.

\begin{figure}[htb]
\begin{center}
\includegraphics[width=13.5cm]{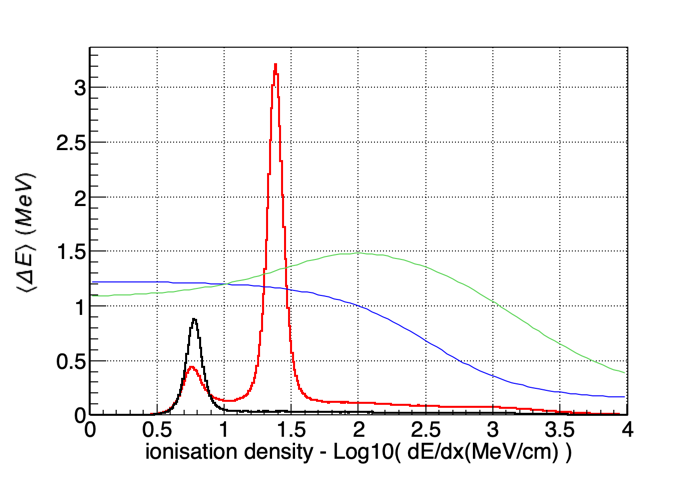}
\end{center}
\caption{Simulated mean ionisation density distribution for Hydrogen nuclei (black) and Helium nuclei (red), at minimum ionization, in a 2-cm thick LYSO layer. Blue and green curves refer to typical relative luminous efficiencies for silicate and alkali scintillator respectively (the curve parameters are that of LYSO and CsI(Tl) in Table \ref{tab:fit}).
\label{fig:MIP_H2_H2}}
\end{figure}

In Figure \ref{fig:MIP_H2_H2} the mean ionisation density distributions produced by $^2$H (black) and $^4$He (red)\footnote{Although not strictly relevant for this study, we simulated A/Z=2 isotopes for consistency with the beam-test data used to tune the model (Section~\ref{sec:beamtest}).} in 2 cm of LYSO is shown. The average is done on about a thousand events with the exclusion of events with nuclear interactions. The secondary peak in the $^4$He distribution at about 6 MeV/cm is due to the $\delta$-ray emission at the highest energies. In the same figure the typical silicate (blue) and alkali (green) relative luminous efficiency in arbitrary unit are shown for comparison.

In case of ideal scintillator materials with a constant luminous efficiency the mean ratio between the light responses would be equal to

\begin{equation}
  R = \frac{\left< S_L(^4He) \right>}{\left< S_L(^2H) \right>} = 4
\end{equation}
as expected by the Z$^2$ ratio between $^4$He and $^2$H. In the real situation $R$ is different from 4 and in particular it will be less or greater than 4 for silicate or alkali scintillator respectively. 

\subsection{Nuclei technique }
\label{sec:technique}

The usual ways  to study and characterise the scintillator non proportionality are the ``electron response'' \cite{Moses1} or the ``photon response'' techniques. 
For example, SLYNCI \cite{SLYNCI} is a facility made especially to characterise the scintillating materials by means of Compton electrons emitted within the materials themselves (Compton Coincidence Technique).

In this work the detection of the ionization produced by high energy nuclei is proposed as an alternative way to investigate the non-proportional effect. The idea to study the scintillator behaviour by detecting high energy nuclei is not new, for example GLAST/FERMI has measured the CsI(Tl) response with beam tests \cite{GLAST} and the DAMPE experiment the BGO response with cosmic rays \cite{DAMPE}. What is proposed here is to characterise quantitatively the light response of different crystals using the theoretical model given in Eq.~(\ref{Model}).

\begin{figure}[htb]
\begin{center}
\includegraphics[width=13.5cm]{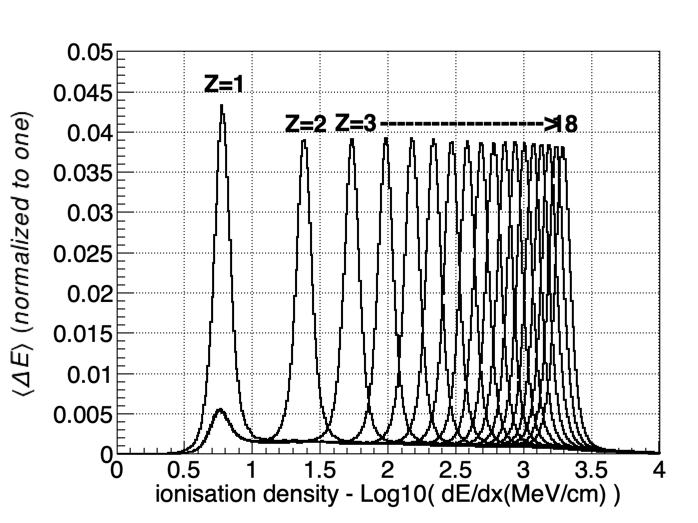}
\end{center}
\caption{Simulated distribution of the mean ionisation density, normalised to one, obtained with nuclei ranging from Hydrogen to Argon, at minimum ionization, in 2 cm of LYSO scintillator.
\label{fig:MIP_LYSOnuclei}}
\end{figure}

As an example of the proposed methodology, the detection of minimum ionising nuclei with Z ranging from 1 to 18 in 2 cm of LYSO is considered. Figure \ref{fig:MIP_LYSOnuclei} shows the mean ionisation density distributions, normalised to one, relative to all the nuclei under study. The eighteen main peaks are spread out over a wide range: this means that they can be used as probes to study the light response as function of the ionisation density between about 5 MeV/cm and 2 GeV/cm.

Let $P_Z$ be the ratio between the mean light signal and the mean energy released in the scintillator for the $Z$-th nucleus:

\begin{equation}
  P_Z \left[ \eta_{e/h}, \left( \frac{dE}{dx} \right)_O, \eta_H, B \right] = \frac{\left< S_L(Z) \right>}{\left< E_{rel}(Z) \right>}
\end{equation}
where the means are estimated by the average value obtained from the simulated events. Only in case of an ideal material $P_Z$ would be independent on $Z$, whereas in the general case it depends on the four parameters of the light emission model in Eq.~(\ref{Model}) through the dependence on $L$. 

The $P_Z$ quantity, obtained by the simulation, is useful to estimate the model parameters by comparing it with the measurements. Let $D_Z$ be the mean signals measured with nuclei divided by $Z^2$. 
For ideal materials and readout systems, $D_Z$ should also be independent on $Z$. In the realistic case, however, that is not the case, but still the dependence of $D_Z$ and $P_Z$ on $Z$ is the same, which allows us, by taking a specific nucleus as reference (e.g. $Z$=18), finding the model parameters, since $P_Z/P_{18} = D_Z/D_{18}$.
(in the following these quantities are called "relative light yield"). In Section \ref{sec:LightResponce} the method of least squares is used to compare the beam test data and the relative luminous efficiency.

\section{ Beam test data }
\label{sec:beamtest}

A comparative test of the response of different scintillator materials was performed in the framework of the CaloCube project \cite{CC0,bongi2015calocube,CC1,vannuccini2017calocube,adriani2016calocube,adriani2017calocube,pacini2017calocube,berti2019calocube, mc,Adriani2019,Adriani2021}. 
CaloCube is an R$\&$D activity   aiming to optimize the design of a wide-acceptance 3-D imaging calorimeter to be operated in space, with the major goal of extending the range of direct CR measurements up to the PeV region.  
The studied configuration consists of a 3D lattice of small cubic scintillating crystals, read-out by photodiodes (PDs), that  
allows an almost isotropic response, so as 
to detect particles arriving from every direction in space. 
The research project included several lines of investigation. The full detector performances were extensively studied by means of a Monte Carlo simulation~\cite{mc}; several tests  were conducted in laboratory aiming to investigate different  
scintillating materials, coatings, photo-sensors, etc. and several prototypes were  constructed~\cite{Adriani2019,Adriani2021}. 

\begin{figure}[htbp]
\begin{center}
\includegraphics[width=13.5cm]{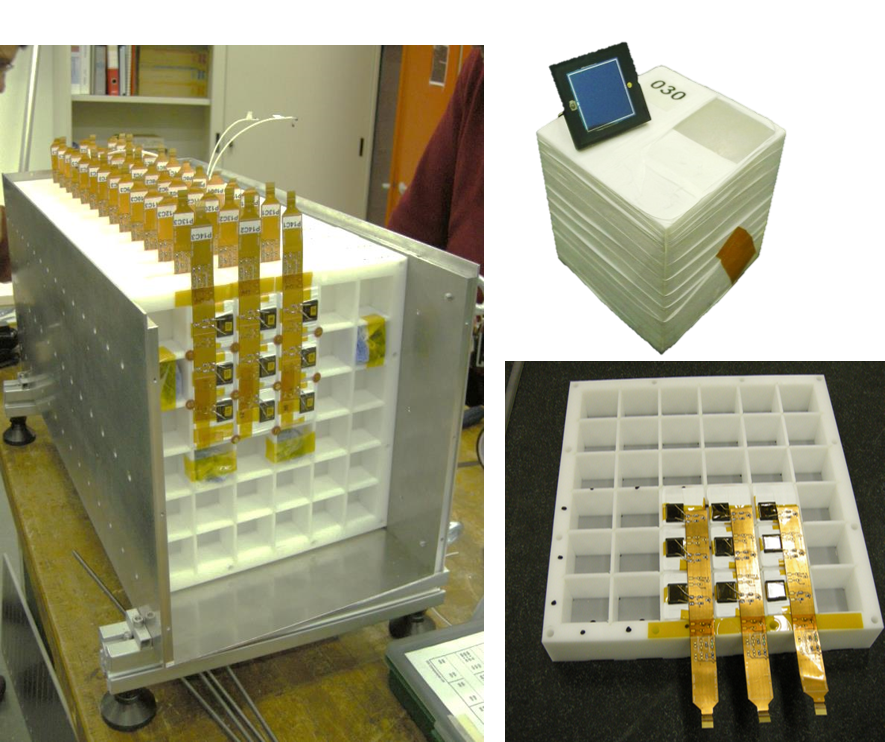}
\end{center}
\caption{ Fully assembled CaloCube prototype  (left) composed of 15 Polyoxymethylene trays. Each tray is loaded with 3$\times$3 CsI(Tl) crystals, wrapped in Teflon and readout by one PD (top right).
The particle beam hits the detector perpendicular to the trays.
\label{fig:pro}}
\end{figure}

In 2015 a medium-scale CsI(Tl) prototype (Fig.~\ref{fig:pro}) was tested with nuclei at the H4 line of CERN Super Proton Synchrotron (SPS). 
The most downstream tray of the prototype was not loaded with CsI(Tl) crystals and was instead used to allocate cubic crystals made of different scintillator materials (we will refer to these scintillators as ``test crystals''). 
During part of the beam test the prototype was rotated 180 degrees so as to place the test crystals in the upstream position.
Two different data sets were collected, 
by exposing the crystals directly to a primary Argon beam of 30$\cdot$A~GeV kinetic energy and to the secondary nuclei beam produced by the Ar fragmentation on a 40-mm thick Polyethylene target.
The magnetic selection system on the beam line allows to choose the A/Z ratio of the secondary products. In our case, we fixed A/Z~2 in order to maximize the beam intensity.

\begin{figure}[htbp]
\begin{center}
\includegraphics[width=6.cm]{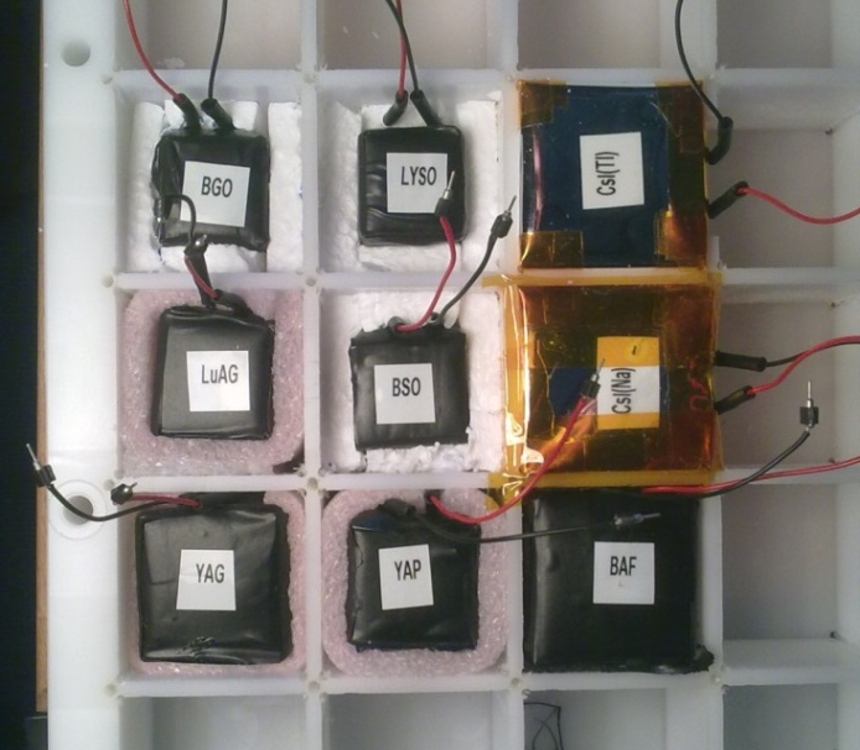} 
\includegraphics[width=7.5cm]{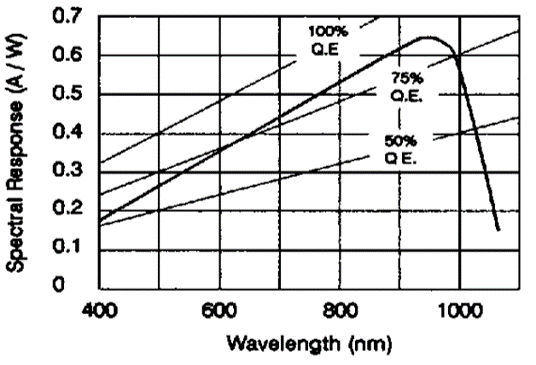}
\end{center}
\caption{ Left: test tray loaded with cubic crystals of different scintillator materials (see Table~\ref{tab:cry}). The particle beam hits the tray perpendicular to its top surface. The photodiode, an  Excelitas VTH2090, is placed on one side of the test crystals.  
Right: spectral response of the photodiode as a function of the wavelength. The straight lines that intersect the curve express the quantum efficiency (Q.E.) of the device, which is maximum at $\approx$~950~nm  and is suitable for all the considered crystals (see Table~\ref{tab:cry}).  
\label{fig:cry}}
\end{figure}

\begin{table}[htbp]
\begin{center}
\begin{tabular}{l|c|c|c|c|c|c}
Material & Size (cm) & $\rho$ (g/cm$^3$) & $\lambda_I$ (cm) & $X_{0}$ (cm) & $\lambda_{max}$ (nm) & $\tau_{decay}$ (ns)\\ \hline
 BGO     & 2.0   & 7.1 & 23 & 1.1 & 480 &  300 \\
 CsI(Tl) & 3.6   & 4.5 & 40 & 1.9 & 550 & 1220 \\
 LYSO    & 2.0   & 7.4 & 21 & 1.1 & 420 &   40 \\
 YAP     & 2.2   & 5.5 & 22 & 2.7 & 370 &   27\\ 
 YAG     & 2.5   & 4.6 & 25 & 3.5 & 550 &  70\\ 
 BaF$_2$ & 3.1   & 4.9 & 31 & 2.0 & 300 &  650\\
 \hline
\end{tabular}
\end{center}
\caption{  List of the tested crystals and main typical characteristics of the scintillator materials. The quantities shown in the table are the length of the side of the cube, the density of the crystal, the interaction and the radiation lengths, the wavelength of the scintillation light at the emission maximum and the decay time of the main emission component.}
\label{tab:cry}
\end{table}



\subsection{Experimental setup}


The same readout system of the CaloCube prototype was used to readout the test crystals. 
An extensive description of the  system can be found in the CaloCube  references (see e.g.  reference~\cite{Adriani2019}). 
We summarize here the main features, relevant for the results discussed in this paper. 

A picture of the test tray is shown in Figure~\ref{fig:cry}. 
Each crystal was coupled to an Excelitas-VTH2090H photodiode having an active area of 84.6~mm$^2$, which has a spectral response suitable for all the considered scintillating crystals (see Figure~\ref{fig:cry} right). 
The PD was placed on one side of the tested crystals, so as to avoid the signal induced by the direct ionization of the photosensor. 
Crystals were wrapped with several layer of Teflon and an outer layer of Tyvek in order to maximize the light collection efficiency.
Out of the nine tested scintillator materials shown in Figure~\ref{fig:cry}, only six were selected for the present study, the choice being based on the relevance of the materials for high-energy applications, the  statistics of the available data set and the signal-to-noise performance of the channel. 
The list of the considered scintillator  materials, with a summary of the main characteristics, is given in Table~\ref{tab:cry}. 

The front-end electronics was based on a high dynamic-range,  low-noise ASIC, specifically designed for calorimetry in space~\cite{CASIS}.   
The main feature of the chip is a double-gain loop, with a gain reduction factor of 1:20 and a real-time selection circuitry:  
if the integrated charge exceeds a certain threshold, the gain is automatically reduced so as to extend the dynamic range of the chip. 
The linearity of the chip is better than 1$\%$ both in high- and low-gain mode~\cite{CASIS}. 
A possible systematic effect of $\approx$~2$\%$  might affect signals acquired in low gain, due to a slight mismatch of the response scale relative to the high-gain regime~\cite{Adriani2021}. 
The noise of the whole system during the beam test was on average 30~ADC~counts, in high gain.
A relevant feature of the chip design is that it is reset periodically, specifically every 20~$\mu$s with the settings applied. 
This implies a variable integration time, depending on the temporal distance between the trigger and the next reset signal, which for slow scintillator materials causes a temporal dependence of the signal. 
The slowest crystal among those considered is CsI(Tl), which also composes the CaloCube calorimeter placed downstream of the test plane. 
In order to minimize the timing effect, only events with integration time larger than 6~$\mu$s were considered.
A time dependent correction was also applied to the signals collected from the CsI(T1) and the  BaF$_2$ test crystals, varying from 0$\%$ to 15$\%$ and 2$\%$, respectively. 
The timing effect is negligible for the other crystals, which have faster decay times. 
Since we are interested in the relative variation of the signal, as the amount of energy deposited varies, no other corrections were applied. Consequently the collected signal represents the convolution among the emission intensity of the crystals, the light-collection efficiency and the quantum efficiency of the PD (Fig.~\ref{fig:cry} right).


A Silicon tracking system~\cite{Marrocchesi:2016gng,Marrocchesi:2014kra} was placed upstream the prototype to provide the particle impact position, with millimeter resolution, and to tag the incident nuclei on the basis of their atomic number Z.   
Only nuclei with a well reconstructed track and traversing the test crystal within its margins were considered.
The CsI(Tl) crystals downstream the test tray were used to select ions that traversed the  test crystal without starting a shower. 
\begin{figure}[t]
\begin{center}
\includegraphics[width=13.5cm]{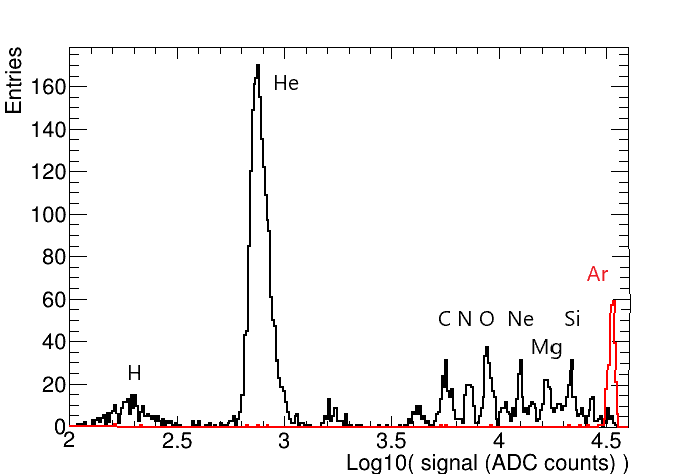}
\end{center}
\caption{ Distributions of the  signals collected by the LYSO  crystal for non-showering nuclei  for the two available data samples, obtained with the Argon beam with (black line) and without (red line) the Polyethylene target.  \label{fig:fra}}
\end{figure}
Figure~\ref{fig:fra} shows, as an example, the  distributions of the signals collected by the LYSO  crystal for the two available data sets, with and without the Polyethylene target, after the selection cuts were applied;  
while the former sample  contains a large variety of nuclei, mostly light, the latter contains mostly Argon. 
As a whole, a large set of elements, spanning from Hydrogen to Argon, was available to study the light response of different scintillator materials as a function of the ionization energy density, as discussed in Section~\ref{sec:technique}.

\begin{figure}[htbp]
\begin{center}
\includegraphics[width=13.5cm]{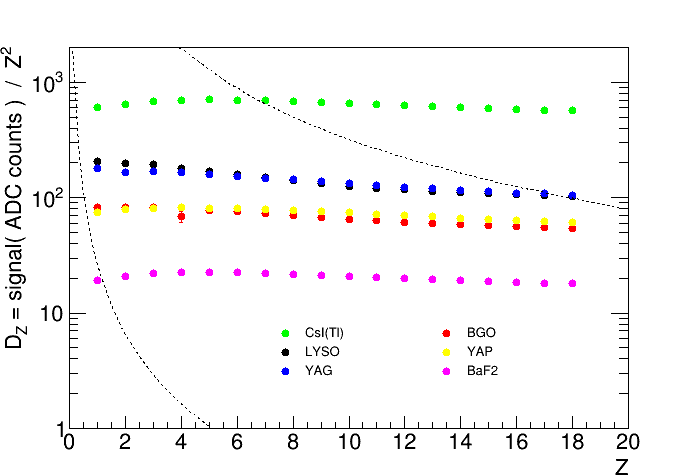}
\end{center}
\caption{ 
Mean value of the signals divided by Z$^2$ ($D_Z$),  collected with the different scintillating crystals for non-showering nuclei, plotted as a function of the atomic number Z. 
The error bars represent the statistical uncertainty. 
The bottom and the top dashed lines represent the noise level (1 sigma) and the threshold between the high- and low-gain regime of the front-end chip, respectively. 
\label{fig:all}}
\end{figure}

A summary of the response of all the tested crystals is shown in Figure~\ref{fig:all}, where the quantity $D_Z$ (see Section~\ref{sec:technique}) is plotted as a function of the atomic number Z of the selected nucleus.   
It is clear from the figure that the light response varies significantly from crystal to crystal. 
As expected, the largest response is obtained with CsI(Tl). 
What emerges from the figure is  that the response of the crystals is not proportional to Z$^2$ and each crystal has its own trend.
The greatest effect of non proportionality appears in the LYSO material with a variation of a factor two ranging from Z=1 to Z=18. LYSO is a recently manufactured type of scintillator, however, the few measurements in the literature support the remarkable non proportionality of the LYSO that we also find \cite{Chewpraditku,Pepin}.
The observed non-proportionality can be explained in terms of the dependence of the luminous efficiency on the ionization density, according to Eq.~(\ref{Model}).

\subsection{Light response model fit \label{sec:LightResponce}}

\begin{figure}[htbp]
\begin{center}
        \begin{minipage}[c]{.49\textwidth}
\centering
          \includegraphics[width=1.0\textwidth]{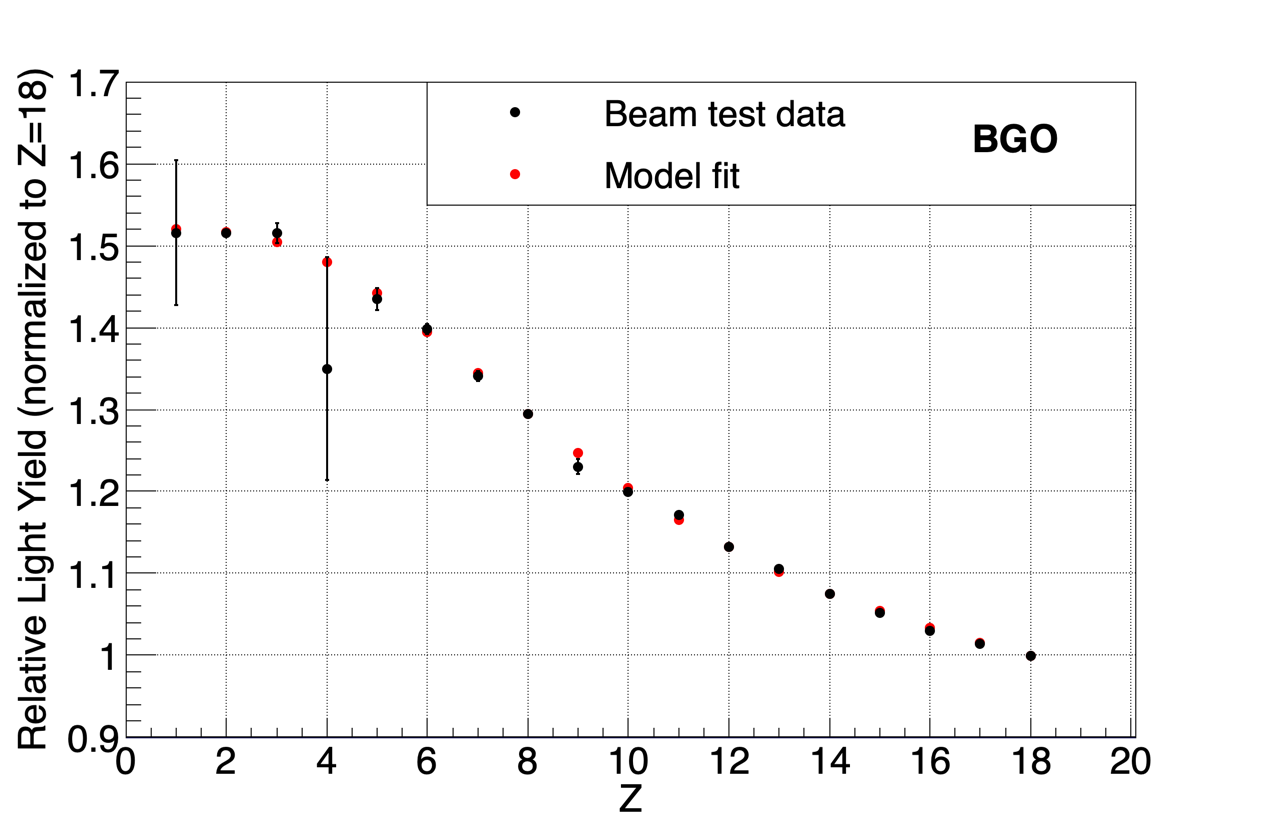}
        \end{minipage}%
        \begin{minipage}[c]{.49\textwidth}
\centering
          \includegraphics[width=1.0\textwidth]{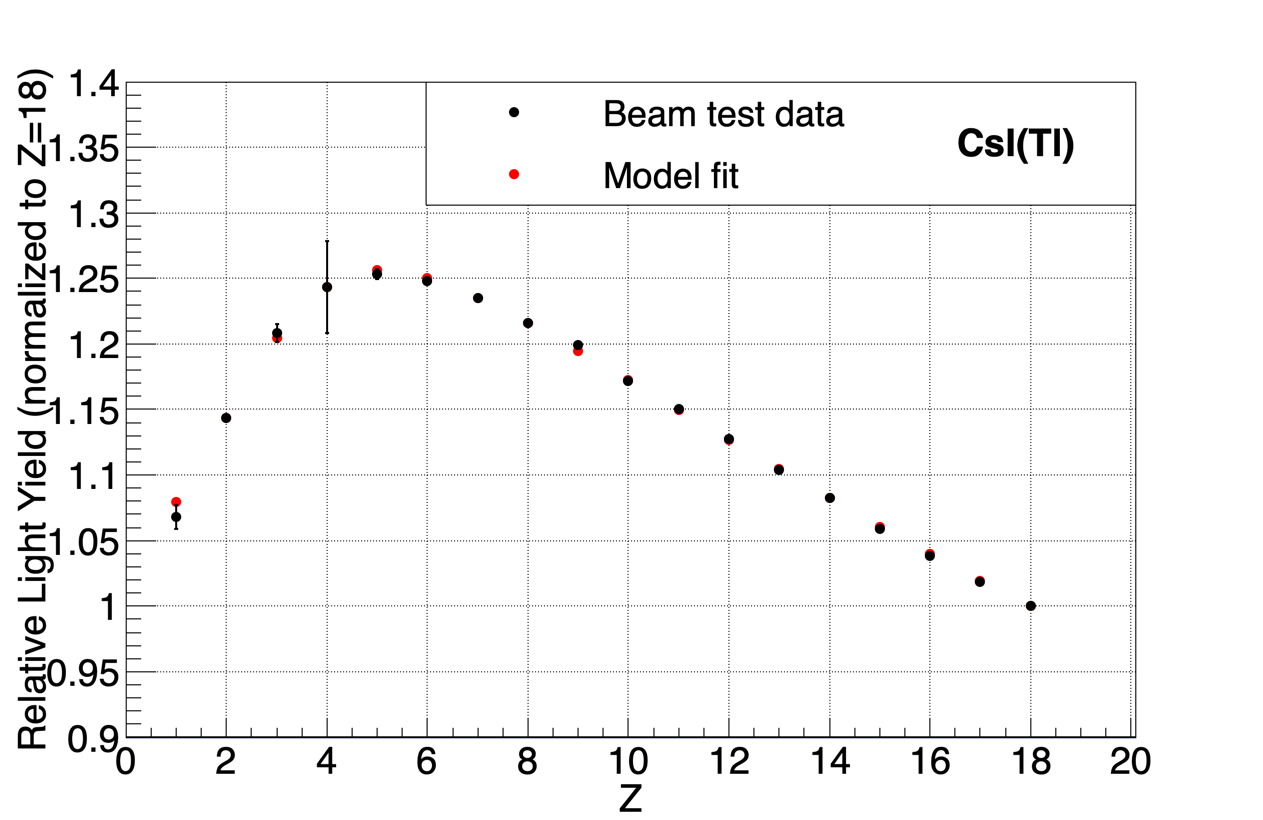}
        \end{minipage}
        \begin{minipage}[c]{.49\textwidth}
\centering
          \includegraphics[width=1.0\textwidth]{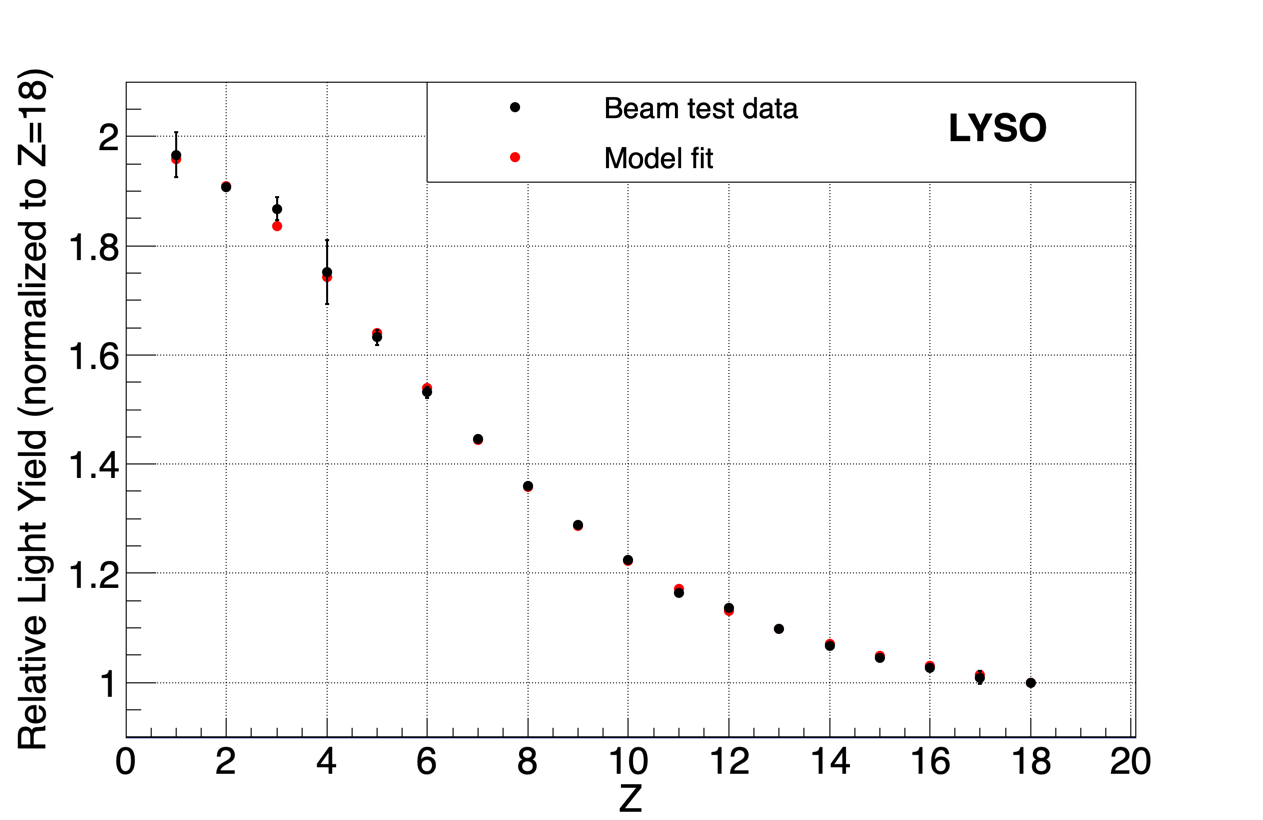}
        \end{minipage}
        \begin{minipage}[c]{.49\textwidth}
\centering
          \includegraphics[width=1.0\textwidth]{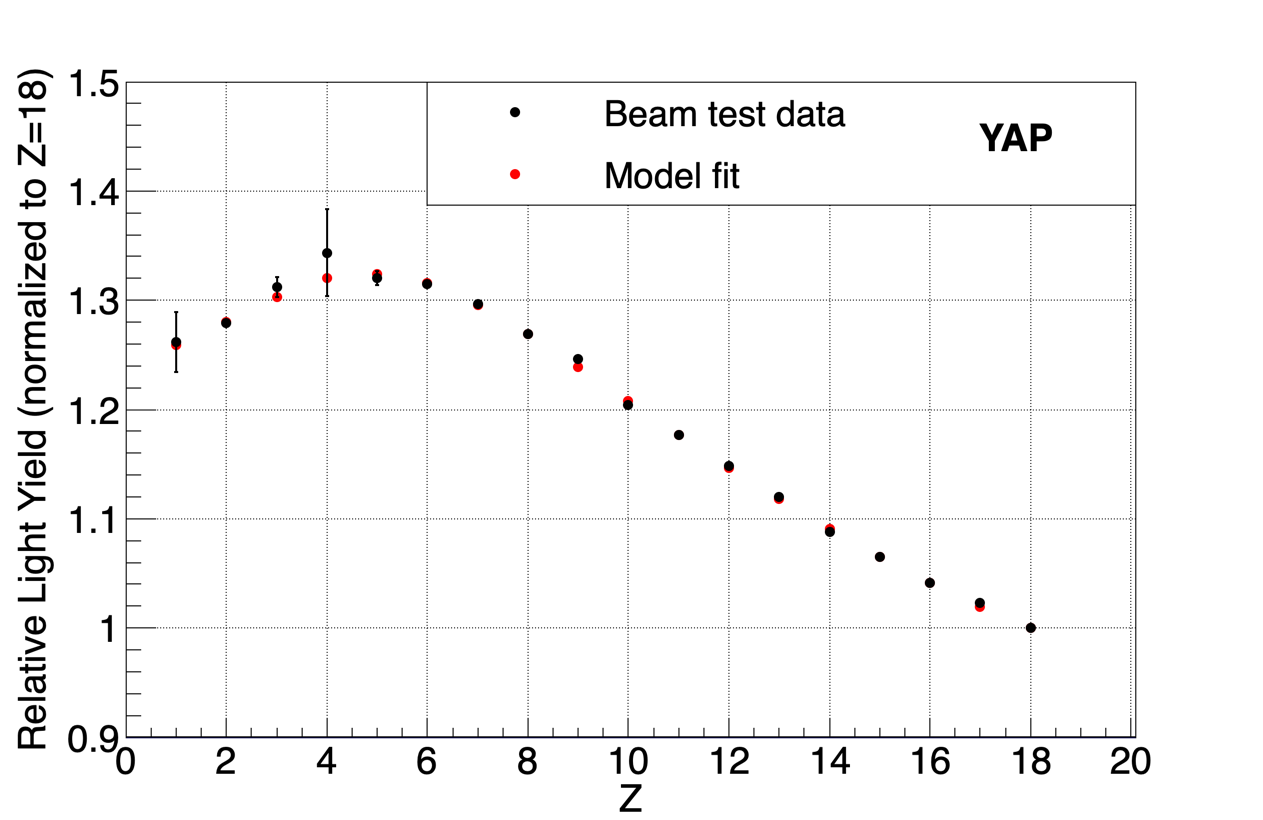}
        \end{minipage}
        \begin{minipage}[c]{.49\textwidth}
\centering
          \includegraphics[width=1.0\textwidth]{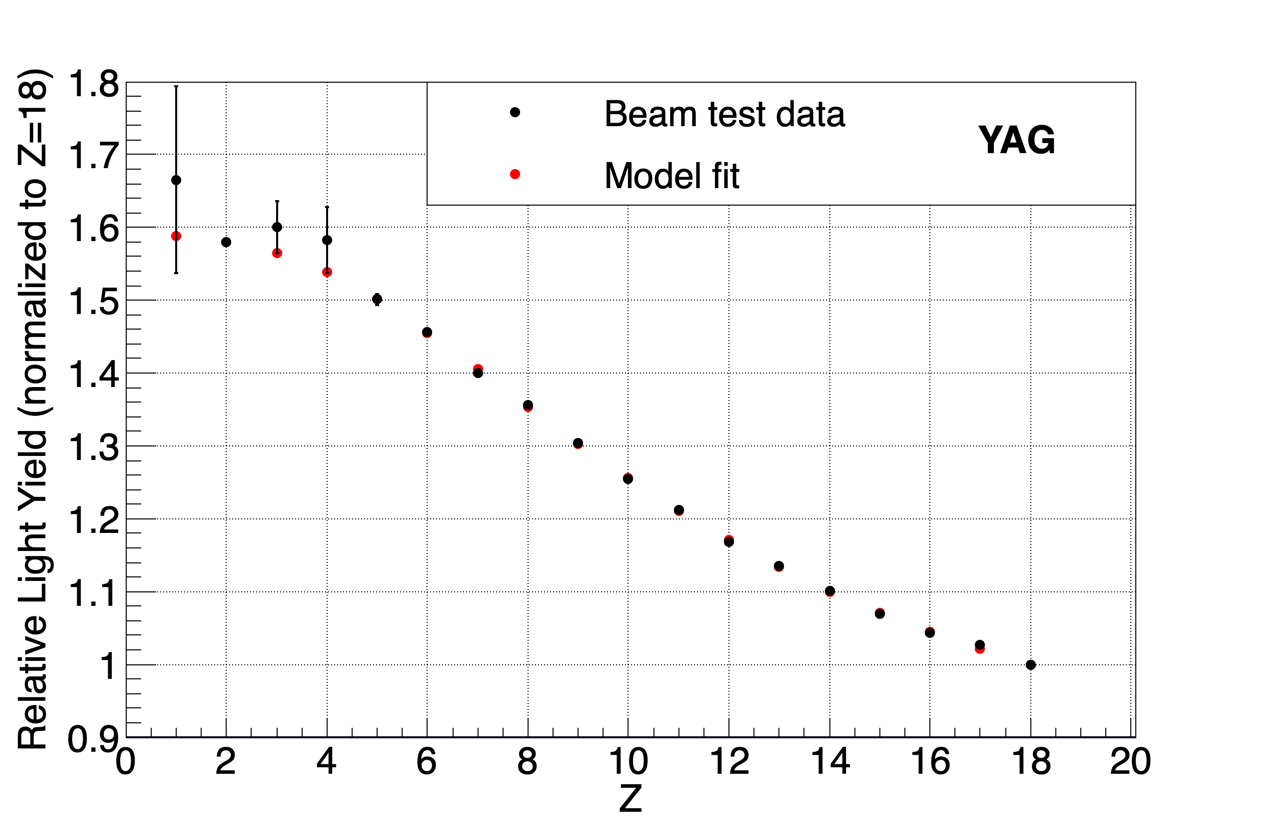}
        \end{minipage}
        \begin{minipage}[c]{.49\textwidth}
\centering
          \includegraphics[width=1.0\textwidth]{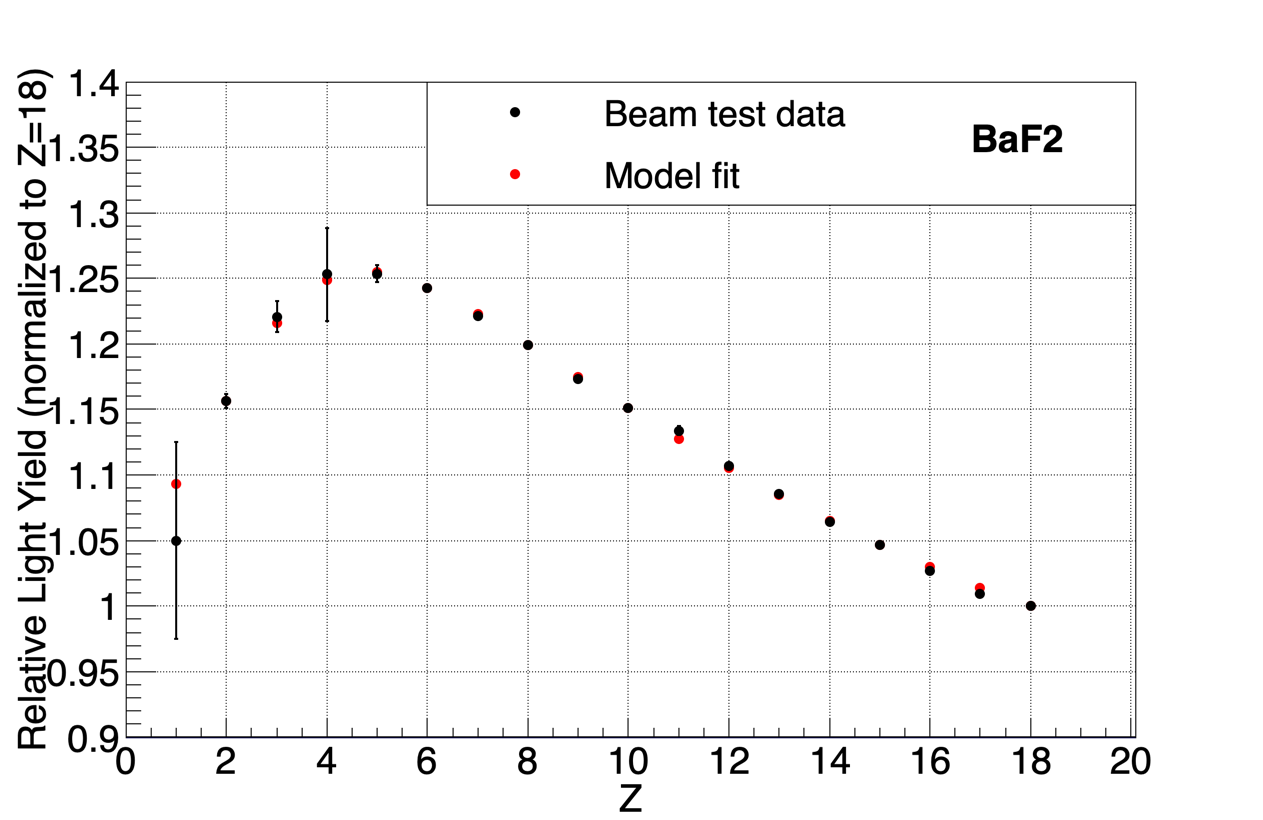}
        \end{minipage}
\end{center} 
\caption{The measured responses of the various test crystals (black points) are compared with the best fit obtained with the "minimalist approach" (red points). 
\label{fig:fit}}
\end{figure}

 The interaction of all nuclei from Hydrogen to Argon with each test crystal has been simulated and the methodology described in Section~\ref{sec:technique} has been used to estimate the model parameters, shown in Table \ref{tab:fit}. 
The interpretation of the parameters in terms of the physical processes invoked to describe the light response of scintillating crystals goes beyond the scope of this work, which is limited to providing a parameterization of the luminous efficiency of different  materials.
 Figure~\ref{fig:fit} shows the good agreement between the measured relative light yield and the model fit results for all the six test crystals. The reduced $\chi^2_{red}$ of the fit procedure, made with the method of least squares, ranges between 0.64 and 1.64. 
 We would like to underline the fact that despite the minimal approach adopted, the model is able to reproduce the experimental trend, which by means of nuclei is measured with excellent precision over a wide range of ionization densities, with great accuracy.

\begin{figure}[htb]
\begin{center}
\includegraphics[width=13.5cm]{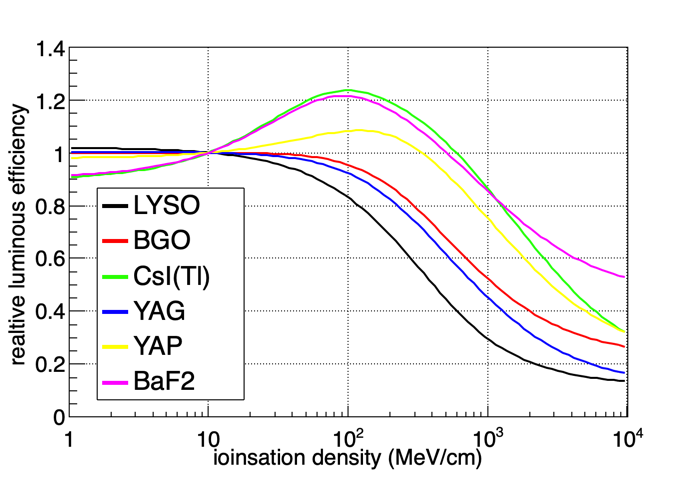}
\end{center}
\caption{Relative luminous efficiency functions obtained from the fit on the measured response of test scintillators, normalised to 1 at 10 MeV/cm.
\label{fig:functions}}
\end{figure}

 The obtained relative luminous efficiency functions are drawn in Figure \ref{fig:functions}, in which the different crystal behaviour of the light response as function of the ionisation density is highlighted.


\begin{table}[htb]
\begin{center}
\begin{tabular}{l|c|c|c|c|c}
Material & $\eta_{e/h}$ & $(dE/dx)_O$ & $\eta_H$ & $(1/B)$ & $p$-value \\ 
   &  & MeV/cm &  & MeV/cm & \\ \hline
 BGO & 0.159 $\pm$ 0.033  &  98 $\pm$ 45 & 0.1884 $\pm$ 0.0039 & 364 $\pm$ 42 & 0.067 \\
 CsI(Tl) & 0.326 $\pm$ 0.010  &  34.1 $\pm$ 2.8 &  0.121 $\pm$ 0.012 & 1338 $\pm$ 64 & 0.65 \\
 LYSO & 0.758 $\pm$ 0.045 &  164.7 $\pm$ 8.4 & 0.0274 $\pm$ 0.0048 & 45.1 $\pm$ 9.1 & 0.82 \\
 YAP & 0.2212 $\pm$ 0.0085 & 90 $\pm$ 11 &  0.174 $\pm$ 0.012 & 873 $\pm$ 70 & 0.24 \\
 YAG &  0.0912 $\pm$ 0.015 & 73 $\pm$ 29 & 0.1052 $\pm$ 0.0055 & 462 $\pm$ 31 & 0.25 \\
 BaF$_2$ & 0.322 $\pm$ 0.024 &  35.8 $\pm$ 6.2 & 0.3440 $\pm$ 0.0071 & 546 $\pm$ 36 & 0.34 \\
\hline
\end{tabular}
\end{center}
\caption{List of the parameter values obtained fitting the signals of non showering nuclei using the "minimalist approach" of the light response.
\label{tab:fit}}
\end{table}

\begin{figure}[htbp]
\begin{center}
          \includegraphics[width=0.95\textwidth]{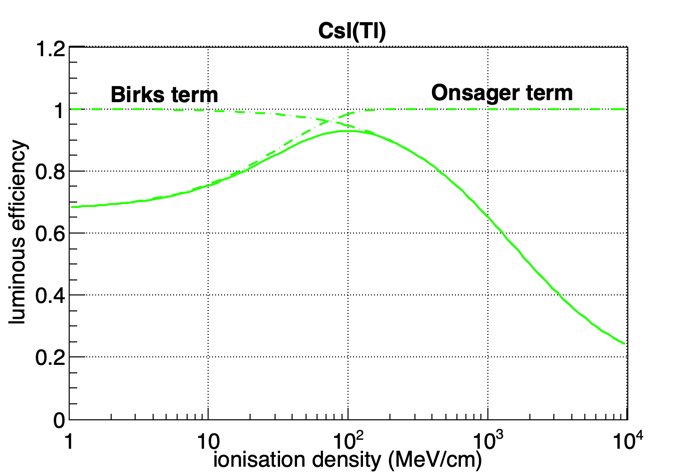}
          \includegraphics[width=1.0\textwidth]{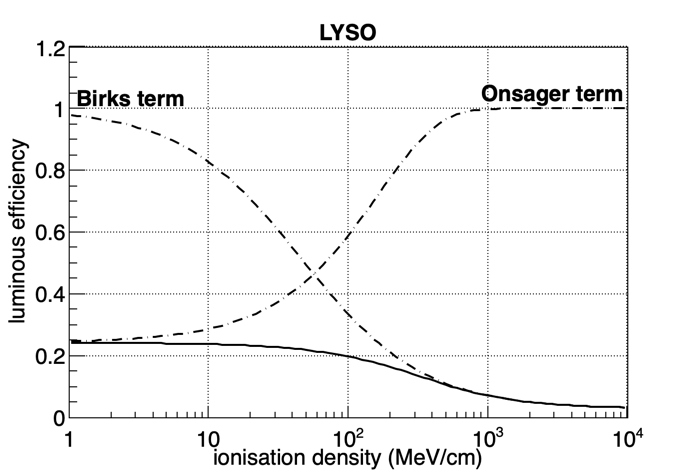}
\end{center} 
\caption{
The dependence of the luminous efficiency in CsI(Tl) and LYSO as a function of the ionisation density showed separately for the Onsager and the Birks effects.}
\label{fig:CsIandLYSO}
\end{figure}

The variety of behaviour of the various materials can be observed. The functional shape of the light response depends on the competing  different strength of the two phenomena considered: the quenching effect (or Birks effect) at high ionisation density and the Onsager effect predominant at low ionisation density. As an example, the dependence of the luminous efficiency for CsI(Tl) and LYSO, separately for the Onsager and the Birks effects, is shown in Figure \ref{fig:CsIandLYSO}. In some cases an increase in luminous efficiency is observed as the ionisation density increases followed by a subsequent decrease. This is the typical case of alkali materials such as CsI or NaI. In other cases the light efficiency curve is monotonically decreasing with the ionization density, as in the case of LYSO. As described in the Section~\ref{sec:showers}, these two different behaviours give rise to opposite sign in the systematic effect on the energy measurements performed with calorimeters. 
\newline
Similar behaviour has been observed through the Compton Coincidence Technique (CCT) with the SLYNCI facility \cite{Payne}. It is worth observing that by means of the CCT it is possible to observe the light emitted by electrons down to an energy of the order of 5 keV which in terms of ionisation density correspond to values of the order of 100 MeV/cm. As shown by Figure~\ref{fig:MIP_LYSOnuclei}, by measuring the nuclei up to Z = 18 a region of ionisation density up to about 2 GeV/cm is explored. 
Despite this, the values obtained in our work for the Onsager term parameter of CsI(Tl), $(dE/dx)_O = 34.1 \pm 2.8$ MeV/cm, appear reasonable when compared with the value 36.4 MeV/cm used in \cite{Payne}. The parameter $\eta_{e/h}$ is also similar: $0.326 \pm 0.010$ in our work and 0.36 in \cite{Payne}.
\newline
The obtained luminous efficiency curves can be dependent on the simulation package used for the ionization processes (the FLUKA code in the present work) and in principle a change in the simulation model can modify the parameters obtained from the "minimalist approach".
\newline
The parameters obtained for a specific material cannot be considered invariant for that material but may vary depending on many factors, for example the crystal manufacturer, as highlighted in \cite{Hull} for the NaI scintillators. Furthermore, the results also depend on the acquisition system used and in particular on the 
signal integration time, because different decay components can have different non-proportionality behaviour (see for example \cite{Syntfeld} for the CsI(Tl) response with different shaping time constants).

\section{ Calorimetric measurements }
\label{sec:showers}

As a final part of this work, possible systematic effects on energy measurements made with calorimeters at high energy are analyzed. Clearly the effects are dependent on the specific material used, on the specific geometry and also on the calibration procedure used. Furthermore, other effects can affect the non-proportional response, as the doping level, the amount of defects and the acquisition system. In order to determine the systematic effect of an experiment it is essential to characterize the specific scintillating material with the same acquisition and electronic system used in the experiment itself. Therefore, the purpose of this analysis is to show the possible existence of systematic effects due to non-proportionality and not to determine quantitatively these effects for one or more specific experiments.

A simple homogeneous cube of 1~m$^3$ of scintillating material is used to perform simulations of the effects on electron and proton shower detection at high energy. The simulated particles are generated at the center and perpendicular to the calorimeter surface. The systematic shifts in the total shower energy measurements depend on the material considered. LYSO, BGO and CsI(Tl) are employed in this study. The MIP value is used as energetic unit for the calorimeter calibration. It is defined as the signal released by high energetic muons (100 GeV) in a thin layer of the same material. In this study the relative light yield can be defined directly as $Y=S_L/E_{rel}$ and a systematic shift in energy measurement is identified when $Y$ is different for minimum ionizing particle (MIP) and shower detection.

\subsection{Electrons}

\begin{figure}[htb]
\begin{center}
\includegraphics[width=13.5cm]{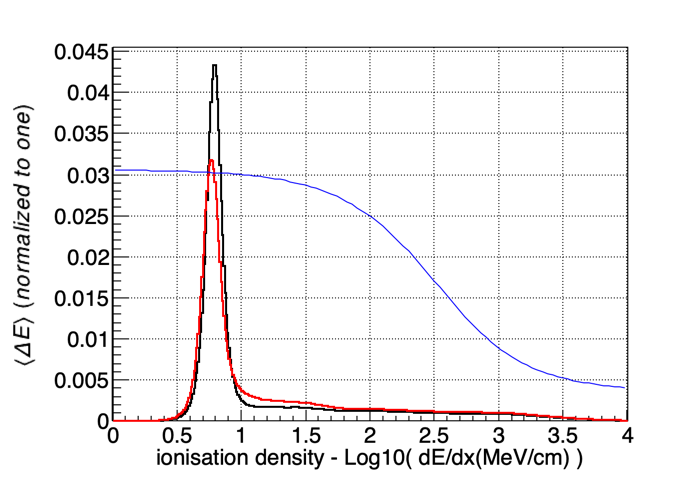}
\end{center}
\caption{Mean ionisation density for 10 GeV electron showers in 1 m depth of LYSO calorimeter (red) and the analogue distribution for muon in a 2 cm layer (black). The blue curve is the relative luminous efficiency function for LYSO.
\label{fig:EleLYSO}}
\end{figure}

In Figure~\ref{fig:EleLYSO} the normalised mean ionisation density distribution for electron showers evaluated at 10 GeV in LYSO is shown. In the figure the electron distribution is compared with the analogous distribution for MIPs in a thin LYSO layer of 2 cm and the estimated relative luminous efficiency function of LYSO is also plotted. The electron distribution is shifted to higher ionisation density values compared to that of the muons, therefore the response is lower when it is weighed with the efficiency function.
 
If we assume that the calorimeter calibration is done using MIP signals, without any correction, the systematic effect of the crystal non-proportionality is given by the relative light yield ratio
\begin{equation}
  \frac{Y(e)}{Y(\mu)} = \frac{S_L(e)}{E_{rel}(e)} \bigg/ \frac{S_L(\mu)}{E_{rel}(\mu)} = 0.977
\end{equation}
which gives rise to an energy measurement shift of about -2.3\%. Simulations made at higher energies (100 GeV and 1 TeV) with 1~m$^3$ of LYSO show that the ionisation density distributions do not change in shape and therefore the systematic effect is independent on the electron energy. 

The ionisation density distribution obtained using CsI or BGO calorimeters are similar to that of the LYSO calorimeter and the systematic effects in the energy measurements are different mainly due to the different luminous efficiencies. The energy measurement shift with 1 m of calorimetric depth are +0.82\% and -1.1\% for CsI(Tl) and BGO, respectively. 

\subsection{Protons}

In this study a depth of 1 m of scintillating material was also used for the study of systematic effects in case of protons showers. It must be kept in mind that while 1 m of calorimetric depth is enough to completely contain an electromagnetic shower this is not true, in general, for high energy hadronic showers. However, the construction of hadronic calorimeters of larger depth, to be used for direct measurements of cosmic rays in space, is currently unfeasible.

\begin{figure}[htb]
\begin{center}
\includegraphics[width=13.5cm]{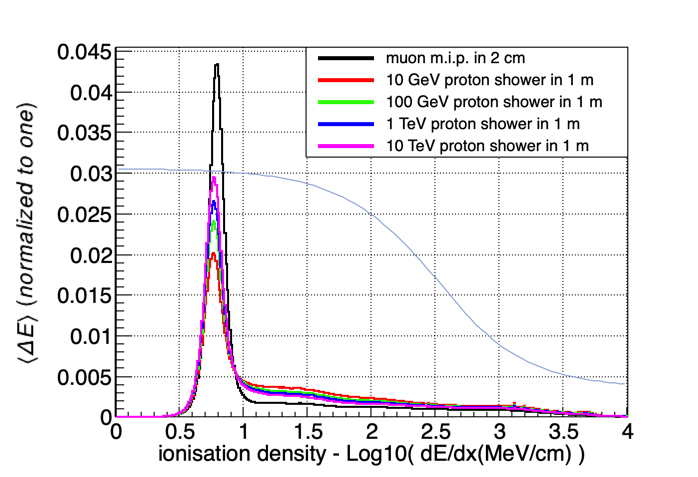}
\end{center}
\caption{Mean ionisation density for proton showers in 1 m depth of LYSO at different energies and the analogue distribution for muon in a 2 cm layer. The curve is the relative luminous efficiency function corresponding to LYSO.   
\label{fig:PrLYSO}}
\end{figure}

In Figure \ref{fig:PrLYSO} the mean ionisation densities relative to proton showers at different energies are plotted (10 GeV, 100 GeV, 1 TeV and 10 TeV). In contrast to electron studies, for protons there is a clear dependence of the density shape as the energy varies. This means that when the total light response of the calorimeter is considered, i.e. the integral of the ionisation density weighted with the luminous efficiency, systematic effects are obtained which depend on energy. In measuring cosmic-ray fluxes, systematic effects that vary with energy imply the introduction of artefacts in the final estimation of the spectrum under observation.

The shape of the ionisation density at the same energies in case of 1 m depth of BGO or CsI(Tl) are analogues to that shown in Figure~\ref{fig:PrLYSO} for LYSO. In Table~\ref{tab:systematics} the calculated systematic effects for electrons and protons in the different materials are summarised. 
The shift in energy for CsI(Tl) has an opposite sign to that for BGO and LYSO due to the different shape of the luminous efficiencies.

\begin{figure}[htbp]
\begin{center}
          \includegraphics[width=0.95\textwidth]{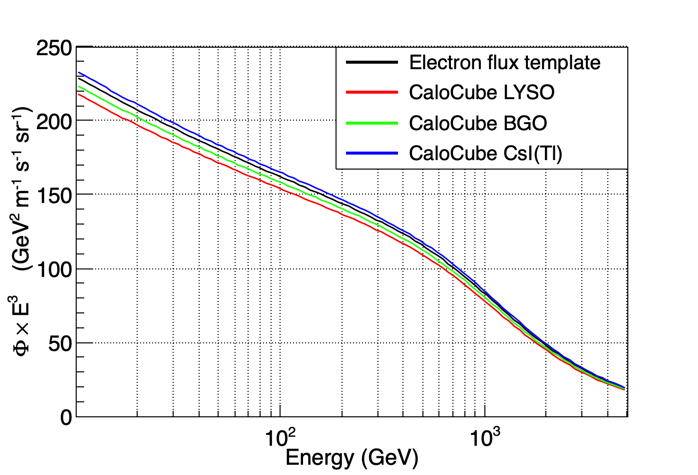}
          \includegraphics[width=1.0\textwidth]{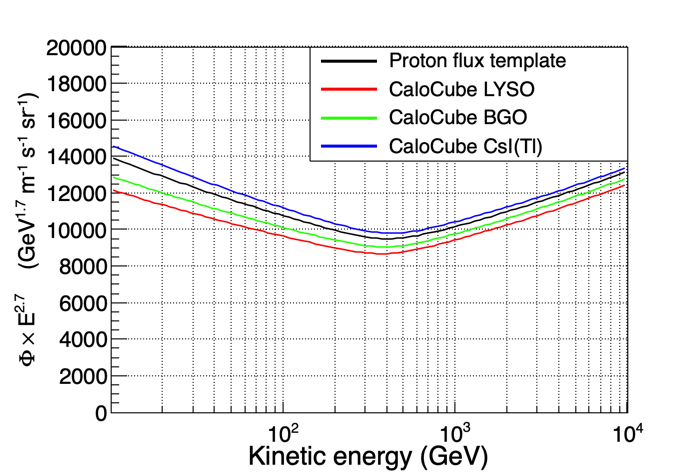}
\end{center} 
\caption{
Measured spectra of electrons and protons affected by the systematic error on the measured energy caused by the  non-proportional light response of inorganic crystals.  
The black line shows the assumed spectral shape, the colored lines show the results obtained using 1~m$^3$ of the considered  scintillator materials (see Table~\ref{tab:systematics}).
\label{fig:systematic}}
\end{figure}

It should be noted that a relative systematic error in the energy measurement equal to $\Delta$ gives rise to a relative  error in the measured  flux equal to $(\gamma-1)\times\Delta$, where -$\gamma$ is the cosmic-ray spectral index.
In Figure \ref{fig:systematic} the expected measured electron and proton spectra in case of the idealized 1 m$^3$ calorimeter used in this simulation work are shown.


\begin{table}[htb]
\begin{center}
\begin{tabular}{l|c|c|c|c|c}
Material & electrons & protons & protons & protons & protons \\ 
   & $\ge$10 GeV & 10 GeV & 100 GeV & 1 TeV & 10 TeV \\ \hline
 LYSO & -2.3\% &  -7.1\% & -5.6\% & -4.6\% & -3.4\% \\
 BGO & -1.1\% &  -4.3\% & -3.0\% & -2.3\% & -1.8\% \\
 CsI(Tl) & +0.82\% &  +2.9\% &  +2.0\% & +1.5\% & +1.2\% \\
 \hline
\end{tabular}
\end{center}
\caption{Systematic shifts in energy measurements obtained simulating 1~m$^3$ cubic calorimeter with electrons and protons at different energies for three materials: LYSO, BGO and CsI(Tl). 
\label{tab:systematics}}
\end{table}

\section{ Conclusions }
\label{sec:conclusion}

Recent direct high-energy measurements of the individual cosmic-ray components, carried out using calorimeters made of scintillating crystals (see Table~\ref{tab:experiments}), have revealed discrepancies among the results that cannot be reconciled with the evaluated errors.
We believe that a deep understanding of the systematic effects affecting this type of measurements is of fundamental importance in view of the future generation of calorimetric experiments, like HERD~\cite{Gargano:2021Q4,Pacini:2021kD}, which aims to carry out precision spectral measurements extended at even higher energy. 
As a first step towards this goal, in this work we pose the problem of how the non-proportional light response typical of scintillating crystals can affect the energy response of the calorimeters and we propose a methodology to evaluate its entity. 

The scintillation light yield depends on the ionization density. 
Beams of high-energy ions, ranging from Hydrogen to Argon, were used to sample the light response of six  different inorganic crystals (BGO, CsI(Tl), LYSO, YAP, YAG and BaF$_2$), by exploiting the different ionization density of the nuclei. 
The measured light signals were modeled considering two major phenomena: the Onsager effect, which acts at low density, and the Birks effect, which acts at high density and gives rise to the well-known quenching effect. 
The resulting luminous efficiencies were used as input of a simulation to evaluate the scintillator response to electromagnetic and hadronic showers. 
In particular, the simulation was used to calculate the distribution of the energy deposited by the secondary particles produced during the development of the shower as a function of the ionization density. 
A comparison of the total light signal collected with different scintillator materials  was then  made, weighting the resulting distribution  with the measured luminous efficiencies.


The result of the calculation indicates that, if the calorimeter response is calibrated by means of minimum ionizing particle signals, a net effect on the total shower signal of the order of few percents exists (see Table~\ref{tab:systematics}). 
The shift can be both positive and negative, depending on the material. 
In particular, if the scintillation light emission is dominated by the Birks effect, the  trend of the luminous efficiency is  monotonically decreasing for increasing ionization density, giving rise to the underestimation of the measured shower energy. 
If, on the other hand, at low energy the Onsager effect is dominant there will first be an increasing trend of the luminous efficiency  followed by a decrease; in this case, depending on the relative strength of the Onsager effect, the net effect can be an overestimation of the shower energy.


The systematic effect resulting from this study could partly explain the discrepancies among cosmic-ray spectra observed by different instruments, even if also other systematic effects, on the energy calibration or on the absolute normalization of the spectrum,  should be investigated.
However, the specific evaluation of the energy shift caused by light yield non-proportionality depends on many factors.  
In fact, as already specified, the luminous efficiency of a scintillator depends on intrinsic characteristics of the crystal, including aspects that depend specifically on the production process, such as impurities.
In addition, the scintillation light response of a real detector also depends on the characteristics of the readout system, such as the quantum efficiency of the photosensor and the signal integration time, since in the case of multiple scintillation components each one can have a different behaviour. 
The correct approach would be to characterize (e.g. by using ion beams as proposed in this work) the same scintillator material with the same readout system used by a specific experiment. 

Finally, besides aspects related to the characteristics of the detector, the evaluation of the energy systematic shift is also affected by the uncertainty on the energy density evaluation, which might significantly depend on the simulation package used (e.g. FLUKA as in this work).





 \bibliographystyle{elsarticle-num} 
 \bibliography{cas-refs}





\end{document}